Marcello Romano

# Exact Analytic Solution for the Rotation of a Rigid Body having Spherical Ellipsoid of Inertia and Subjected to a Constant Torque

**Abstract** The exact analytic solution is introduced for the rotational motion of a rigid body having three equal principal moments of inertia and subjected to an external torque vector which is constant for an observer fixed with the body, and to arbitrary initial angular velocity. In the paper a parametrization of the rotation by three complex numbers is used. In particular, the rows of the rotation matrix are seen as elements of the unit sphere and projected, by stereographic projection, onto points on the complex plane. In this representation, the kinematic differential equation reduces to an equation of Riccati type, which is solved through appropriate choices of substitutions, thereby yielding an analytic solution in terms of confluent hypergeometric functions. The rotation matrix is recovered from the three complex rotation variables

The author is with the
Mechanical and Astronautical Engineering Department, and with the Space Systems Academic Group, Naval Postgraduate School, 700 Dyer Road, Monterey, California
Tel.: ++1-831-656-2885
Fax: ++1-831-656-2238
E-mail: mromano@nps.edu

by inverse stereographic map. The results of a numerical experiment confirming the exactness of the analytic solution are reported. The newly found analytic solution is valid for any motion time length and rotation amplitude. The present paper adds a further element to the small set of special cases for which an exact solution of the rotational motion of a rigid body exists.

**Keywords** rigid body dynamics, kinematics, rotation, integrable cases of motion, spherical ellipsoid of inertia

**1 Introduction**

The rotational motion of a rigid body has been studied for centuries. Besides the theoretical interest, the subject is of practical importance for many applications, pertaining, in particular, to astronautics and celestial mechanics.

The problem of the rotational motion of a rigid body can be divided into two parts. The *dynamic problem* aims to obtain the angular velocity of the rigid body with respect to an inertial reference by starting from the knowledge of the initial angular velocity and the history of the applied torque. On the other hand, the *kinematic problem* focuses on determining the current orientation of the body from the knowledge of the initial orientation and the history of the angular velocity. The differential equations governing the rotational dynamic and kinematic problems are decoupled if the applied torque does not depend on the body orientation.

The exact analytic solution for the rotational motion of a rigid body exists only for a small number of special cases.

When no torque acts on the body, exact solutions exist for both the dynamic and kinematic problems in the case of asymmetric rigid body with a fixed point coincident with the center of mass (Euler-Poinsot case), and in the one of axis-symmetric rigid body under gravity force with a fixed point along the axis of symmetry (Lagrange-Poisson case) (Golubev 1960; Leimanis 1965; Morton et al. 1974).

On the contrary, when the rigid body is subjected to external torques, the exact analytic solution exists, limited to the the dynamic problem, only for an axis-symmetric rigid body under constant axial torque (Bödewadt 1952; Longuski 1991). Nevertheless, no exact analytic solution has been known for the kinematic problem, with the exception of the simple case of torque along a principal axis of inertia and initial angular velocity along the same axis.

Several researchers have proposed approximate solutions. In particular, Longuski 1991, and Tsiotras and Longuski 1991, and Longuski and Tsiotras 1995 use asymptotic and series expansion to get the approximate solution of the dynamics and kinematics for both the constant and the time-varying torque cases. Livneh and Wie 1997 analyze qualitatively, in the phase space, the dynamics of a rigid-body subjected to constant torque. As regards the kinematic problem, Iserles and Nørsett 1999 study the solution in terms of series expansion for the more general problem of solving linear differential equation in Lie Groups, building upon the work of Magnus 1954. Finally, Celledoni and Saefstroem 2006 propose ad-hoc numerical integration algorithms.

For the first time, to the knowledge of the author, the present paper introduces the exact analytic solution for the rotational kinematic problem of a rigid body with a spherical ellipsoid of inertia subjected to a torque vector which is constant with respect to an observer fixed with the body, and to arbitrary initial angular velocity. Besides a uniform sphere, any uniform regular simplex has spherical ellipsoid of inertia (Klamkin, 1993). The proposed analytic solution is valid for any length of time and rotation amplitude, as opposed to the previously known approximate solutions cited above.

In particular, the parametrization of the rotation matrix in terms of complex stereographic rotation variables, introduced earlier by Tsiotras and Longuski 1991, is used in the paper in order to obtain a kinematic differential equation of Riccati type which is then solved analytically by substitutions of variables. The solution involves the confluent hypergeometric function.

The proposed analytic solution is of considerable theoretical interest, as it adds a new element to the set of few special cases for which an exact solution of the rotational motion of a rigid body exists. Therefore, it can be used to rapidly obtain accurate solutions over arbitrary large intervals of time, and constitutes a significant new comparison case for error analysis of approximate algorithms.

The paper is organized as follows: Section 2 introduces the dynamic and kinematic equations for the rotational motion of a generic rigid body, and the parametrization of the rotation by stereographic projection, Section 3 reports the equations for the particular case of a rigid body having spherical ellipsoid of inertia and subjected to a constant torque, Section 4 introduces the new analytic integration of those equations. Finally, Section 5 presents the results of the numerical experiment, and Section 6 concludes the paper.

## 2 General Rigid-Body Problem Statement

For a generic rigid body the Euler's rotational equations of motion are (Goldstein 1980)

$$
\begin{aligned}
I_1 \dot{p} &= (I_2 - I_3)\, q\, r + m_1 \\
I_2 \dot{q} &= (I_3 - I_1)\, r\, p + m_2 \\
I_3 \dot{r} &= (I_1 - I_2)\, p\, q + m_3,
\end{aligned}
\qquad (1)
$$

where the dot symbol denotes the time derivative, $\{I_i : i = 1, 2, 3\}$ are the moments of inertia about the three orthogonal body-fixed principal axes of inertia, $\{p, q, r\}$ are the components along those axes of the angular velocity of the body with respect to the inertial frame and $\{m_i : i = 1, 2, 3\}$ are the components along the same axes of the external torque acting on the body.

The rotation transformation matrix $R_{NB} \in SO(3)$ from a body fixed coordinate system $B$ to the inertial coordinate system $N$ is defined such that

$$ a_N = R_{NB}\, a_B, \qquad (2) $$

where $a_N$ is the column matrix containing the components of a vector **a** in the Cartesian coordinate system $N$, and $a_B$ is the column matrix of components of the same vector in the Cartesian coordinate system $B$.

The matrix $R_{NB}$ obeys the following differential equation (Leimanis 1965)

$$\dot{R}_{NB} = R_{NB}\Omega, \qquad (3)$$

where

$$\Omega = \begin{bmatrix} 0 & -r & q \\ r & 0 & -p \\ -q & p & 0 \end{bmatrix}, \qquad (4)$$

Eq. 3 is the kinematic differential equation for the rigid body. Integration of this equation provides the current body orientation, and, together with the integrals of Eq. 1, completely solves the rigid body motion problem.

In general, Equations 1 and 3 do not have exact analytic solutions.

2.1 Kinematic differential equations in terms of complex stereographic variables

Besides the rotation transformation matrix, several other set of variables can be used to describe the orientation, or attitude, of a rigid body (Shuster 1993).

In particular, the set of three complex stereographic rotation variables proposed in Longuski and Tsiotras 1995 is used in the present paper. The definition of these variables is summarized here below.

Because $R_{NB}$ is orthonormal, the elements of its k-th row ($r_{k1}$, $r_{k2}$ and $r_{k3}$) are the coordinates of a point on the unit sphere $\mathcal{S}^2$ in $R^3$.

The stereographic projection from the south pole of the unit sphere onto the equatorial complex plane establishes a map between $\mathcal{S}^2$ and the extended complex plane $C \bigcup \infty$.

In particular, the three points $\{(r_{k1}, r_{k2}, r_{k3}) : k = 1, 2, 3\}$ are projected onto the following three complex numbers (Longuski and Tsiotras 1995)

$$w_k = \frac{r_{k2} - i\, r_{k1}}{1 + r_{k3}}, \quad k = 1, 2, 3, \tag{5}$$

where $i$ denotes the imaginary unit. These constitute the stereographic rotation variables.

Given the stereographic rotation variables, the elements of the matrix $R_{NB}$ can be found by exploiting the inverse stereographic map from the complex plane onto the unit sphere. In particular, it results (Longuski and Tsiotras 1995)

$$r_{k1} = \frac{i\,(w_k - \overline{w_k})}{1 + |w_k|^2}, \quad r_{k2} = \frac{w_k + \overline{w_k}}{1 + |w_k|^2}, \quad r_{k3} = \frac{1 - |w_k|^2}{1 + |w_k|^2}, \quad k = 1, 2, 3, \tag{6}$$

where $|\bullet|$ denotes the absolute value of a complex number and the over bar indicates the complex conjugate.

By differentiating Eq. 5 and by using Eq. 3, after some algebraic steps, the rotational kinematic differential equations in terms of the stereographic rotation variables $w_k$ are found to be

$$\dot{w}_k = \frac{1}{2}\,(p - i\,q)\,w_k^2 - i\,r\,w_k + \frac{1}{2}\,(p + i\,q), \quad k = 1, 2, 3. \tag{7}$$

These equations, which are of the Riccati form, do not have in general an exact analytic solution.

**3 Case of a rigid body with spherical ellipsoid of inertia and constant torque**

For a rigid body having a spherical ellipsoid of inertia, i.e. either a uniform sphere or any uniform regular simplex (Klamkin, 1993),

$$I_1 = I_2 = I_3 = I. \tag{8}$$

In this case, Eq. 1 become

$$\dot{p} = m'_1$$
$$\dot{q} = m'_2 \qquad (9)$$
$$\dot{r} = m'_3,$$

where $\{m'_i = m_i/I : i = 1, 2, 3\}$.

Let us now assume that the torque acting on the body is constant with respect to a body-fixed reference frame.

A suitable rotational transformation of coordinates can be used to align the third axis of a new body-fixed coordinate system along the direction of the acting torque. Therefore, without loss of generality, Eq. 9 can be written as

$$\dot{p} = 0$$
$$\dot{q} = 0 \qquad (10)$$
$$\dot{r} = U,$$

where $p, q$ and $r$ are used, from this point on, to indicate the components of the angular velocity along the new reference axes, and U indicates the constant torque, normalized by the value of the moment of inertia.

The integration of Eq. 10 immediately gives

$$p(t) = p_0$$
$$q(t) = q_0 \qquad (11)$$
$$r(t) = r_0 + Ut,$$

where $p_0, q_0$ and $r_0$ denote the initial condition values, and $t$ denotes the time.

Even if the solution of the dynamic problem has the simple linear form of Eq. 11, the corresponding kinematic differential equation, which is obtained by substituting the relations in Eq. 11 into Eq. 4 and 3, cannot be directly integrated, with the exception of when the initial angular velocity is along the same axis of the acting torque ($p_0 = q_0 = 0$). In this case the solution is

$$R_{NB}(t) = R_{NB}(0) \exp\left(\int_0^t \Omega(\xi)\, d\xi\right), \qquad (12)$$

where $\exp(\bullet)$ indicates the matrix exponential. Eq. 12 is only valid if the matrix $\Omega(t)$ commutes with its integral (Adrianova 1995). This commutativity is lost when the initial angular velocity is arbitrary; therefore, Eq. 12 is not valid anymore in that case. The limitation in the validity of the solution of Eq. 12 is not clearly stated in Leimanis 1965, as already noticed in Longuski 1984.

In order to find the analytic solution for an arbitrary initial angular velocity, we consider the kinematic differential equations in terms of the stereographic variables. In particular, by substituting the relations in Eq. 11 into Eq. 7, it results

$$\dot{w}(t) = \frac{1}{2}(p_0 - iq_0)\, w(t)^2 - i\,(r_0 + Ut)\, w(t) + \frac{1}{2}(p_0 + iq_0) \qquad (13)$$

where the subindex $k$ has been dropped in order to unburden the notation.

## 4 Exact analytic integration of the kinematic differential equation

**Theorem 1** *Given $p_0, q_0, r_0$ and $U$ real numbers, with $p_0$ and $q_0$ not both zero, the general solution of Eq. 13, governing the rotational kinematics of a rigid body with spherical ellipsoid of inertia, initial angular velocity components $p(0)=p_0$, $q(0)=q_0$, and $r(0)=r_0$ along the three body fixed axes $\sigma_1, \sigma_2$ and $\sigma_3$, and subjected to a constant torque $U$, normalized by the value of the moment of inertia and directed along the axis $\sigma_3$, is the following*

$$w(t) = \frac{(1+i)\sqrt{U}}{3\,(p_0 - iq_0)}\,[6z + G(z)], \qquad (14)$$

*with*

$$G(z) := \frac{2\,{}_1\mathrm{F}_1\!\left(\frac{3-\nu}{2}, \frac{5}{2}; z^2\right)(\nu - 1)\,z^2 + 6\,{}_1\mathrm{F}_1\!\left(1 - \frac{\nu}{2}, \frac{3}{2}; z^2\right)c\nu z - 3\,{}_1\mathrm{F}_1\!\left(\frac{1-\nu}{2}, \frac{3}{2}; z^2\right)}{{}_1\mathrm{F}_1\!\left(-\frac{\nu}{2}, \frac{1}{2}; z^2\right)c + {}_1\mathrm{F}_1\!\left(\frac{1-\nu}{2}, \frac{3}{2}; z^2\right)z}, \qquad (15)$$

where $_1F_1$ denotes the confluent hypergeometric function (Lebedev 1965), $c \in C$ is the constant of integration and

$$z := \frac{(1+i)(r_0 + Ut)}{2\sqrt{U}}, \qquad \nu := -1 - \frac{i(p_0^2 + q_0^2)}{4U}. \qquad (16)$$

*Proof* Three successive variable substitutions are used in this proof. In the first substitution, the dependent variable in Eq. 13 is changed from $w(t)$ to $x(t)$ by the Riccati transformation (Zwillinger 1989)

$$w(t) = -\frac{2\dot{x}(t)}{x(t)(p_0 - iq_0)}, \qquad (17)$$

then the equivalent second order homogeneous linear differential equation is obtained

$$\ddot{x}(t) + i(r_0 + Ut)\dot{x}(t) + s_0 x(t) = 0, \qquad (18)$$

where $s_0 := (p_0^2 + q_0^2)/4$.

In the second substitution, the dependent variable in Eq. 18 is changed from $x(t)$ to $y(t)$ by

$$x(t) = y(t) \exp\left[-it\left(r_0 + \frac{Ut}{2}\right)\right], \qquad (19)$$

then the following equivalent equation is obtained

$$\ddot{y}(t) - i(r_0 + Ut)\dot{y}(t) + (s_0 - iU)y(t) = 0. \qquad (20)$$

Finally, the independent and dependent variables in Eq. 20 are changed, from $t$ to $z$, as defined in Eq.16, and from $y(t)$ to $\eta(z) = y(t)$ respectively. Therefore, the following equivalent equation is obtained

$$\frac{d^2\eta(z)}{dz^2} - 2z\frac{d\eta(z)}{dz} + 2\nu\,\eta(z) = 0, \qquad (21)$$

where $\nu$ is defined as in Eq. 16. This is a second order differential equation of Hermite type (Zwillinger 1989) and has the following general solution (Lebedev 1965)

$$\eta(z) = {}_1F_1\left(-\frac{\nu}{2}, \frac{1}{2}; z^2\right) a + {}_1F_1\left(\frac{1-\nu}{2}, \frac{3}{2}; z^2\right) b\,z, \qquad (22)$$

with $a \in C$ and $b \in C$ constants of integration.

By substituting the expression in Eq. 22 back into Equations 19 and 17, by defining $c = a/b$ and by rearranging, the result stated in Eq. 14 and 16 is finally obtained.

**Corollary 1** *An alternative expression of the solution given in Eq. 14 and 16 is*

$$w(t) = \frac{2(1+i)\sqrt{U}}{p_0 - iq_0} \left\{ z - \frac{\nu \left[ H_{\nu-1}(z)\, c' - {}_1F_1\left(1 - \frac{\nu}{2}, \frac{3}{2}; z^2\right) z \right]}{H_\nu(z)\, c' + {}_1F_1\left(-\frac{\nu}{2}, \frac{1}{2}; z^2\right)} \right\}, \quad (23)$$

*where* H *denotes the Hermite function (Lebedev 1965), $c' \in C$ is the constant of integration, and $z$ and $\nu$ are defined as in Eq.16.*

*Proof* Since the Hermite function is by definition a particular solution of the Hermite equation, we can write the general solution of Eq. 21 as

$$\eta(z) = {}_1F_1\left(-\frac{\nu}{2}, \frac{1}{2}; z^2\right) a' + H_\nu(z)\, b', \quad (24)$$

with $a' \in C$ and $b' \in C$ constants of integration.

By using the expression in Eq. 24 instead of the one in Eq. 22, by defining $c' = a'/b'$, and by proceeding as for the proof of Theorem 1, the result in Eq. 23 is obtained.

The confluent hypergeometric function used in Theorem 1, which is defined in $C^3$, is (Lebedev 1965)

$$ {}_1F_1(\alpha, \beta; \gamma) = \sum_{k=0}^{\infty} \frac{(\alpha)_k}{(\beta)_k} \frac{\gamma^k}{k!}, \quad |\gamma| < \infty, \quad \beta \neq 0, -1, -2, ..., \quad (25)$$

where $(\bullet)_k$ denotes the Pochammer symbol

$$(\bullet)_0 = 1, \quad (\bullet)_k = k(\bullet + 1)\ldots(\bullet + k - 1), \quad k = 1, 2, \ldots. \quad (26)$$

The series in Eq. 25 converges for all finite $\gamma$, and it represents an entire function of $\gamma$, for fixed $\alpha$ and $\beta$.

4.1 Solution in terms of rotation matrix

The problem of computing the rotation matrix function $R_{NB}(t)$ is solved by finding the three complex functions $w_1(t), w_2(t)$ and $w_3(t)$, particular integrals of Eq. 13 which substituted into Eq. 6 give the elements of the first, second and third row, respectively, of the rotation matrix.

Without loss of generality, we can consider the rotation matrix $R_{NB}$ to be equal to the identity matrix at the initial time $t = 0$. Therefore, from Eq. 6, the initial conditions for $w_1(t), w_2(t)$ and $w_3(t)$ are found to be

$$
\begin{aligned}
(r_{11} = 1, r_{12} = 0, r_{13} = 0) &\Rightarrow w_1(0) = -i \\
(r_{21} = 0, r_{22} = 1, r_{23} = 0) &\Rightarrow w_2(0) = 1 \\
(r_{31} = 0, r_{32} = 0, r_{33} = 1) &\Rightarrow w_3(0) = 0.
\end{aligned}
\tag{27}
$$

The values $\{c_1, c_2, c_3\}$ of the constants of integration, which substituted in the general integral (Eq. 14) give the three particular integrals $w_1(t), w_2(t)$ and $w_3(t)$, are found by solving Eq. 14 for $c$ and by imposing the three initial conditions of Eq. 27

$$
\begin{aligned}
c_1 &= -\frac{(1+i)}{6\sqrt{U}} \left[ \frac{{}_1F_1\left(\frac{1-\nu}{2}, \frac{3}{2}; \frac{ir_0^2}{2U}\right)\left(6r_0^2 + 3(p_0 - iq_0)r_0 + 6iU\right) + 2g}{2\,{}_1F_1\left(1-\frac{\nu}{2}, \frac{3}{2}; \frac{ir_0^2}{2U}\right)\nu r_0 + {}_1F_1\left(-\frac{\nu}{2}, \frac{1}{2}; \frac{ir_0^2}{2U}\right)(p_0 - iq_0 + 2r_0)} \right] \\
c_2 &= \frac{(i-1)}{6\sqrt{U}} \left[ \frac{{}_1F_1\left(\frac{1-\nu}{2}, \frac{3}{2}; \frac{ir_0^2}{2U}\right)\left(6ir_0^2 - 3(p_0 - iq_0)r_0 - 6U\right) + 2gi}{2\,{}_1F_1\left(1-\frac{\nu}{2}, \frac{3}{2}; \frac{ir_0^2}{2U}\right)\nu r_0 + {}_1F_1\left(-\frac{\nu}{2}, \frac{1}{2}; \frac{ir_0^2}{2U}\right)(ip_0 + q_0 + 2r_0)} \right] \\
c_3 &= -\frac{(1+i)}{6r_0\sqrt{U}} \left[ \frac{g + 3\,{}_1F_1\left(\frac{1-\nu}{2}, \frac{3}{2}; \frac{ir_0^2}{2U}\right)\left(r_0^2 + iU\right)}{{}_1F_1\left(1-\frac{\nu}{2}, \frac{3}{2}; \frac{ir_0^2}{2U}\right)\nu + {}_1F_1\left(-\frac{\nu}{2}, \frac{1}{2}; \frac{ir_0^2}{2U}\right)} \right],
\end{aligned}
\tag{28}
$$

where $\nu$ is defined as in Eq. 16, and

$$
g := {}_1F_1\left(\frac{3-\nu}{2}, \frac{5}{2}; \frac{ir_0^2}{2U}\right)(\nu - 1)r_0^2.
\tag{29}
$$

An alternative method to compute the rotation matrix consists of using the above procedure to compute the elements of two rows of the matrix only, and then exploit the orthonormality conditions in order to find the elements of the remaining row. In particular, this method allows to overcome the problem of the point at infinite of the stereographic projection and the related singularity of the inverse stereographic map. In fact, whenever one of the body axes approaches the direction of the south pole of the unit sphere and, consequently, the correspondent point, mapped by the stereographic projection (see Eq. 5), tends to infinite, it is sufficient to consider only the remaining two axes in order to build the rotation matrix. This procedure makes the proposed analytic solution valid for any length of time and rotation amplitude.

## 5 Numerical experiments

A numerical experiment has been performed by using the newly found solution and its results are reported here below. The experiment considered a rigid body having spherical ellipsoid of inertia and moving with the following sample initial conditions and body fixed torque value

$$p_0 = 10, \quad q_0 = 15, \quad r_0 = 20, \quad U = 3. \tag{30}$$

The numerical evaluation of the analytic solution is compared to the solution obtained by a numerical integration method at a sample particular time of $t = 40$.

The numerical evaluation of the exact analytic solution in Eq. 14, with the constants of integration evaluated as in Eq. 28, gives the following values for the stereographic rotation variables

$$\begin{aligned} w_1 &= -0.4263512625196356 + 0.4033450082324547\,i \\ w_2 &= -0.2092049992603608 - 0.5057428050364536\,i \\ w_3 &= 0.4167055594784583 + 0.1095946611492525\,i. \end{aligned} \tag{31}$$

By substituting the above values into the relations in Eq. 6, the corresponding rotation matrix results

$$\tag{32}$$

$$R_{NB} = \begin{bmatrix} -0.6000092673712773 & -0.6342329852754623 & 0.4875832231087923 \\ 0.7783397597095152 & -0.3219671485837583 & 0.5390031295717849 \\ -0.1848677838995137 & 0.7029122815980806 & 0.6868320222985118 \end{bmatrix}.$$

The software package Mathematica was used to conduct the above numerical evaluations. This computation took 0.8 seconds of CPU time.

For validation purposes, a direct numerical integration of Eq. 3 was conducted by using the Matlab algorithm ode113 (Adams-Bashforth-Moulton), with relative and absolute tolerances set to $10^{-14}$. This computation took 11 seconds of CPU time. The numerical evaluation of the analytic solution given in Eq. 32 and the result of the numerical integration coincides up to the $11^{th}$ digits. This relative error between the two computations is attributed to the accuracy of the numerical integrator being lower than the one of the numerical evaluation of the analytic solution. This is confirmed by the corresponding different level of accuracy in maintaining the orthonormality of the rotation matrix.

## 6 Conclusions

For the first time, to the knowledge of the author, the exact analytic solution has been introduced in this paper for the rotational motion of a rigid body having spherical ellipsoid of inertia (e.g. an homogeneous cube or sphere, or a sphere made of homogeneous concentric layers) subjected to a constant torque and arbitrary initial angular velocity. Therefore, the present paper adds a new significant element to the set of few special cases for which an exact solution of the rotational motion of a rigid body exists. Furthermore, the proposed analytic solution is valid for any length of time and rotation amplitude.

In particular, the parametrization of the rotation by three complex stereographic variables was used in order to obtain a kinematic differential equation of Riccati type which was then solved analytically by substitutions of variables. The solution involves confluent hypergeometric functions. The rotation matrix can be recovered from the complex rotation variables by inverse stereographic map.

The results of the numerical experiment clearly confirms the exactness of the proposed analytic solution.

**Acknowledgements**

The author would like to thank the editor and the anonymous reviewers for their useful comments. This paper is dedicated to Paola, Alice and Matilde.